\documentstyle[prl,aps]{revtex}
\large
\begin{document}
\twocolumn
\draft
\title
{Toulouse limit for the overscreened four-channel Kondo problem}

\author{M. Fabrizio$^{(a)}$ and Alexander O. Gogolin$^{(a,b)}$}

\address{
$^{(a)}$ Institut Laue--Langevin,
BP 156, 38042 Grenoble, France\\
$^{(b)}$ Landau Institute for Theoretical Physics,
Kosygina str. 2, Moscow, 117940, Russia} \maketitle

%\begin{center}
%\end{center}

\begin{abstract}
We show that the spin dynamics of the 4--channel Kondo model
is equivalent to the one of a model with a spin 1/2 impurity
coupled to spin 1 conduction electrons. By abelian bosonization
of the latter model we find a kind of Toulouse limit analogous to
that recently discussed by Emery and Kivelson for the 2--channel case.
We analyse the model by exploiting the close analogy to the
problem of an impurity in the Luttinger liquid.
\end{abstract}
\smallskip
\bigskip
\pacs {}
\narrowtext

The overscreened multi--channel Kondo model
has been the subject of an intensive analysis after
Nozi\`eres and Blandin's discovery of
its non--Fermi--liquid behavior\cite{boss}.
Various methods have been used
to tackle this problem: Bethe ansatz technique\cite{Bethe},
large $n$ (the number of channels) expansion\cite{boss,LargeN}
and conformal field theory\cite{Tsvelik,AL}.
The main feature of the overscreened Kondo model is a
non--analytic temperature and magnetic field dependence
of the impurity correction to the free energy, which manifests
in the fractional exponents characterizing
the power law behavior of, e.g., susceptibility and specific heat.
The exact exponents and the asymptotic behavior of the dynamic  correlation
functions have been determined
by means of the Bethe ansatz and conformal field theory.

Recently, Emery and Kivelson found an alternative approach
to the problem based on simple abelian bosonization\cite{EK}.
It is analogous to the well known Toulouse limit for the
standard single channel Kondo model\cite{Toulouse} and allows one to
re--obtain  all the exact results plus additional information on dynamic
properties. However their method is only applicable to the case $n=2$.

In this paper we show that a similar approach can be devised for
the four channel model. The corresponding Toulouse limit
is not exactly solvable on the contrary to the $n=1,2$ case.
Nevertheless, we are able to determine its low energy properties,
which are sufficient to identify all the exponents
which control the low temperature behavior of the thermodynamic quantities.

The hamiltonian for the $n$--channel Kondo model,
$H=H_0+H_{exc}$,
consists of the term describing free electrons,
\[
H_0 = v_F\sum_{i=1}^{n}\sum_{\sigma=\uparrow,\downarrow}
 \int_{-\infty}^{+\infty} dx \psi^\dagger_{i\sigma} (x)
(-i\partial_x )\psi^{\phantom{\dagger}}_{i\sigma}(x) \;
\]
and the exchange term,
\[
H_{exc}=
\sum_{a=x,y,z} I_a S_a J^a (0)\; ,
\]
where $I_a$ are the exchange constants, $S_a$ are the impurity
spin--$1/2$ operators, and $J^a (x)$ are the electron
spin currents (densities):
\begin{equation}
J^a (x)=\sum_i \psi^\dagger_{i\sigma} (x) \tau^a_{\sigma\sigma'}
\psi^{\phantom{\dagger}}_{i\sigma'}(x)\; ,
\label{currents}
\end{equation}
$\tau^a$ being spin 1/2 matrices.
As usually (see, e.g., \cite{AL}), we have retained
the s--wave scattering only, linearized the fermion spectrum,
and replaced the outgoing and ingoing waves with $n$--copies of
right--moving electron fields
$\psi^{\phantom{\dagger}}_{i\sigma}(x)$ defined for
$-\infty < x < +\infty$. Following \cite{AL} we introduce charge and ``flavor"
currents,
\[
J(x)=\sum_{i\sigma} \psi^\dagger_{i\sigma}
(x)\psi^{\phantom{\dagger}}_{i\sigma}(x),\;
J^A (x) = \sum_{ij\sigma}
\psi^\dagger_{i\sigma} (x) t^{A}_{ij}\psi^{\phantom{\dagger}}_{j\sigma}(x)
\]
$t^{A}_{ij}$ being generators of $SU(n)$ group, and
re--write the free part of the hamiltonian as a sum of
three commuting terms:
\[
H_0=\frac{v_F}{2\pi}
 \int dx \left[ \frac{J(x) J(x) }{4n} +\frac{J^A(x)  J^A(x)  }{n+2}
+ \frac{\vec{J}(x) \vec{J}(x) }{n+2}  \right]
\]
This allows one to formulate the problem entirely in
terms of the electron spin
currents $\vec{J}(x)$. The information about the number
of channels is contained
in the commutation relations obeyed by these currents
[which follow from their definition Eq.(\ref{currents})]:
\begin{equation}
[J^a (x), J^b (0)]= i\epsilon^{abc}\delta(x) J^c (0)- i\frac{n}{4\pi}
\delta^{ab} \delta' (x)
\label{KacMoody}
\end{equation}
indicating that $J^a (x)$ form $SU(2)_{n}$
Kac--Moody algebra (see, e.g., \cite{G}). In order to study
magnetic properties, we include also a magnetic field $h$, which gives rise
to an additional term in the hamiltonian
\[
h\left[ S_z + \int dx J_z(x)\right].
\]

Our idea is to model the commutation relations (\ref{KacMoody}),
for particular values of $n$, by some free
fermion representation of the currents, which will eventually be
more convenient
to work with (in a sense to be specified later)
then the original representation given by Eq.(\ref{currents}).

The following currents:
\begin{eqnarray}
J^{+} (x) &=&
 \sqrt{2}\left[ \psi^\dagger_0 (x)\psi^{\phantom{\dagger}}_{-1}(x)
+ \psi^\dagger_1
(x)\psi^{\phantom{\dagger}}_{0}(x)\right],\nonumber\\[-0.1truecm]
&~ &~~~~~~ \label{currentsbis} \\[-0.1truecm]
J^z (x) &=& \psi^\dagger_1 (x)\psi^{\phantom{\dagger}}_{1}(x)
- \psi^\dagger_{-1} (x)\psi^{\phantom{\dagger}}_{-1}(x),\nonumber
\end{eqnarray}
where $\psi^{\phantom{\dagger}}_{m}(x)$, $m=-1,0,1$, are spinless
right--moving free
electron fields,
can be shown to obey the commutation relations (\ref{KacMoody})
with $n=4$.
One can think of Eq.(\ref{currentsbis}) as a representation of
spin 1 electron currents.

Since the spin dynamics is completely determined by the commutation relations
obeyed by the electron spin currents, we can equally well work with
one or the other set of $J^a$'s, Eq.~(\ref{currents}) or (\ref{currentsbis});
the spin properties should be the same. This has also an interesting
outcome that the spin behavior of the four channel Kondo model
is equivalent to the one of a spin 1/2 impurity coupled to spin 1
electrons. The latter problem is interesting by itself, being the simplest
realization of a Kondo model with integer spin conduction electrons.

What we are going to show is that the representation (\ref{currentsbis})
has the advantage of having a simple Toulouse limit, which
the original representation (\ref{currents}) seems not to exhibit.
We bosonize the fermion fields according to
\[
\psi^{\phantom{\dagger}}_{m}(x)=(2\pi\alpha)^{-1/2} {\rm e}^{i\phi_m (x)},
\]
where $\phi_m (x)$ are free right-moving Bose fields and $\alpha$ stands for
a high--energy cutoff (lattice spacing).
Combining these Bose fields as
\begin{eqnarray*}
\phi(x) &=& [\phi_1(x)+\phi_0(x)+\phi_{-1}(x)]/\sqrt{3}, \\
\phi_s(x) &=& [\phi_1(x)-\phi_{-1}(x)]/\sqrt{2},\\
\phi_f(x) &=& [\phi_1(x)-2\phi_0(x)+\phi_{-1}(x)]/\sqrt{6},
\end{eqnarray*}
we see that the field $\phi(x)$ decouples from the impurity, and the
part of the hamiltonian, $H_K$, which is relevant for the Kondo effect,
takes the form \cite{noteA}:
\begin{eqnarray}
H_K &=& H_0(\phi_s)+H_0(\phi_f) +
\frac{\sqrt{2}I_\perp}{\pi\alpha}\left[
S_x \cos\left(\frac{\phi_s}{\sqrt{2}}\right)-\right.\nonumber\\
& &\left. S_y\sin\left(\frac{\phi_s}{\sqrt{2}}\right)\right]
\cos\left(\sqrt{\frac{3}{2}}\phi_f\right) + \label{hambose}\\
& &\frac{I_z}{\sqrt{2}\pi} S_z \partial_x \phi_s
+ h \left[ S_z + \int dx \frac{1}{\sqrt{2}\pi}\partial_x
\phi_s(x)\right],\nonumber
\end{eqnarray}
where $H_0(\phi)$ is the free Bose field hamiltonian,
\[
H_0 (\phi )=
\frac{v_F}{4\pi} \int dx [ \partial_x \phi (x) ]^{2}
\]
and $I_x =I_y =I_\perp$ is assumed (for convenience
everywhere we do not indicate explicitly the $x$--dependence of the Bose
fields,
we intend them at the impurity site $x=0$).

The next step is to employ the canonical transformation:
\begin{equation}
H_K\to UH_KU^\dagger ,\;\;\;
U=\exp\{ - i S_z \phi_s / \sqrt{2} \}.
\label{canonical}
\end{equation}
The transformed hamiltonian reads:
\begin{eqnarray}
H_K &=& H_0(\phi_s )+H_0(\phi_f) +
\frac{\sqrt{2}I_\perp}{\pi\alpha}
S_x \cos\left(\sqrt{\frac{3}{2}}\phi_f\right)+ \nonumber \\
& &\delta H + \frac{h}{\sqrt{2}\pi}\int dx \partial_x \phi_s(x),
\label{hamlambda}
\end{eqnarray}
with
\begin{equation}
\delta H =\lambda S_z \partial_x \phi_s,
\label{deltaH}
\end{equation}
where $\sqrt{2}\pi\lambda =I_z -\pi v_F$. In order to get rid of the
bulk contribution of the magnetic field, we shift the Bose field
according to
\[
\partial_x \phi_s(x)\to \partial_x \phi_s(x) - \sqrt{2}h/v_F \; .
\]
This shift changes
\begin{equation}
\delta H \to \lambda S_z ( \partial_x \phi_s- \sqrt{2}h/v_F ),
\label{deltaHnew}
\end{equation}
and generates a constant $-1/2 \chi_0 h^2$ times the length of the system,
where $\chi_0=1/\pi v_F$ is the uniform spin susceptibility of
free conduction electrons.

We see that the line $\lambda =0$ is indeed analogous to the Toulouse limit
for the single--channel Kondo problem and to the Emery--Kivelson line for the
two--channel one. Along this line, the canonical transformation
(\ref{canonical}) leads to the decoupling of the impurity degrees of freedom
from the conduction electron ones
[the impurity spin component $S_x$ commutes with the hamiltonian
(\ref{hamlambda}) and hence
loses its dynamics]. However, the hamiltonian for the phase field $\phi_f$
remains nontrivial. Let us write it in the form:
\begin{equation}
H(\phi_f )= H_0(\phi_f ) \pm I_\perp
\cos\left(\sqrt{2g}\phi_f\right)/\sqrt{2}\pi\alpha\; ,
\label{hamphi}
\end{equation}
where for convention the $\pm$ sign refers to the conserved
spin component $S_x$ equal $\pm 1/2$.
In our case $g=3/4$. For the two--channel problem, the same hamiltonian was
found with $g=1/2$ in Ref.~\cite{EK}.

We observe that the hamiltonian (\ref{hamphi}) is equivalent to the one
describing a backscattering of electrons from a single impurity in the spinless
Luttinger Liquid.
The hamiltonian for this model, expressed in terms of the
original Fermi operators, is:
\[
H= H_{bulk} + I_\perp
 \left( \Psi^\dagger_R(0)\Psi^{\phantom{\dagger}}_L(0) + H.c.
\right)/\sqrt{2}
\]
where the field $\Psi_{R(L)}$ refers to
right (left) moving fermions, and $H_{bulk}$ is the interacting
Luttinger liquid hamiltonian. If one bosonizes,
according to the standard rules, these Fermi fields
and gets rid of the interaction by performing a Bogoliubov rotation,
the resulting hamiltonian is indeed Eq.~(\ref{hamphi})\cite{KF}.
The parameter $g$ in that equation is a measure of the strength
of the fermion--fermion interaction, being $g=1$ for non--interacting
fermions and $g<1(g>1)$ for repulsive(attractive) interaction.
The problem of an impurity in the Luttinger liquid
recently became quite popular in the context of the physics
of one--dimensional quantum wires, and it was studied in great
detail\cite{KF}.
We will make use of some known results for this problem to
analyse the, as we showed, equivalent Kondo model.
Since in the specific case we are considering the parameter $g=3/4<1$,
the impurity scattering operator is relevant and $I_\perp$ flows to
infinity under scaling transformation. The resulting (stable)
fixed point describes
a perfectly reflecting barrier, which cuts the system into two
semi--infinite lines $a$ and $b$. Exactly at the fixed point,
that is for everything regarding zero energy and temperature
properties, the two regions $a$ and $b$ are disconnected.
At any finite frequency there is a residual
tunneling across the barrier, described by the operator
\begin{equation}
O=\Psi^\dagger_a(0)\Psi^{\phantom{\dagger}}_b(0) + H.c.\;,
\label{tunnel}
\end{equation}
where $\Psi_{a(b)}$ is
the electron field referred to region $a(b)$. This operator scales to zero
approaching the fixed point as $\omega^{1/g -1}$,
with correlation function $\langle O(t)O(0)\rangle\sim t^{-2/g}$
(see Ref.~\cite{KF}).

Coming back to the Kondo model (\ref{hamlambda}) and
(\ref{deltaH}),
we notice that on the line $\lambda=0$ the
total uniform susceptibility $\chi$ coincides with $\chi_0$,
that one of the conduction electrons in the absence of the magnetic impurity.
Thus the impurity susceptibility defined as $\chi_{imp}=\chi-\chi_0$
vanishes on this line (the same occurs at the equivalent line
of two--channel Kondo model\cite{AL,EK}).
Similarly the
impurity contribution to the total specific heat at $\lambda=0$ vanishes.
Therefore, in order to study the physical properties of the Kondo model
we need to perform
a perturbation expansion in the correction $\delta H$, Eq.~(\ref{deltaHnew}),
to the hamiltonian (\ref{hamlambda})
around the solvable line $\lambda=0$ \cite{EK}. As a first step, we calculate
the dimension $x$ of the impurity spin $S_z$, which determines
the large time decay of the correlation function
\begin{equation}
\langle T\left(S_z(t)S_z(0)\right)\rangle = C t^{-2x}\; ,
\label{SzSz}
\end{equation}
where $C$ is a numerical coefficient.

Since $S_z$ is a sum of lowering and raising operators
for the states labeled by the $S_x$ component of the spin,
its effect is simply to reverse the sign of the backscattering
potential $I_\perp$ in the hamiltonian
(\ref{hamphi}).
Therefore the problem of finding the dimension of $S_z$ is similar, but
not exactly equivalent, to the problem of calculating
the X--ray edge exponents in the Luttinger liquid. The latter problem
corresponds to a sudden switching of the backscattering potential on(off)
and has recently been studied\cite{Xray}. The difference
in our case is that, instead of switching on(off) the scattering
potential, we must cope with the sudden change of its sign.
For non--interacting electrons ($g=1$) the dimension of this operator
can easily be calculated by a standard approach based on phase--shift
arguments \cite{schon}, and it turns out to be $x=1/4$.
For the value $g=1/2$ the exponent $x=1/2$ was recently found
in Ref.~\cite{EK}. For arbitrary $g\leq 1$, we find
$x=1/4g$. This result, although related to the x--ray edge singularity,
differs in a striking way, specifically the maximum value of the
exponent is not bounded by the scattering phase shift value
in the unitary limit.

The exponent $x=1/4g$ is calculated by noticing that the unitary operator
$W=\exp \{ i \pi J_t/2 \}$
changes sign of the backscattering potential provided that the
operator $J_t$ is the total fermionic current\cite{notab}, which is
defined as the difference between the total numbers of right and
left moving fermions and satisfies
\[
[J_t,\phi_f(x)]=i\sqrt{2/g} \; .
\]
Therefore
\[
\langle T\left( S_z(t)S_z(0)\right)\rangle \propto
\langle T\left( W(t)W(0)\right)\rangle .
\]
Due to the backscattering potential, the current $J_t$ is not
a conserved quantity but acquires a
dynamics which has been analysed in Ref.\cite{GP}.
{}From this analysis, $J_t$ turns out to be essentially a gaussian
variable with correlation function behaving at large time as
\[
\langle T\left( J_t(t)J_t(0)\right)\rangle =
(2/\pi^2 g ) \ln t.
\]
Hence the dimension of $W$ (i.e. $S_z$) is indeed $x=1/4g$.
The dimension of the operator $\delta H$, Eq.~(\ref{deltaH}),
is $1+1/4g>1$, thus
being irrelevant at the stable zero temperature
fixed point $\lambda=0$, $I_\perp=\infty$. The finite temperature
behavior of the system is described by the irrelevant operators
which move the hamiltonian away from this fixed point, specifically
the operator $\delta H$, which brings the system away from $\lambda=0$,
and the operator $\tilde{\lambda}O$ [see Eq.~(\ref{tunnel})],
corresponding to a deviation from $I_\perp=\infty$, where
$\tilde{\lambda}$ is a non--universal parameter.
As we showed, the former operator has dimension $1+1/4g$ and the latter one
has dimension $1/g$.
For our case, $g=3/4$, the two dimensions coincide,
which implies that both operators have to be considered on equal footing.
The two--channel Kondo model corresponds to $g=1/2$, then
$O$ is more irrelevant than $\delta H$ and can be neglected.

Armed with this results, we can evaluate the first non--vanishing
correction to the free energy due to a finite $\lambda$ and
$\tilde{\lambda}$ in the presence of a magnetic field.
The contribution proportional to $\lambda^2$ is
\begin{eqnarray*}
& &\delta F^{(\lambda)}_{imp}
= -\frac{1}{2} \int_{0}^{\beta} d\tau \langle T [
\delta H (\tau) \delta H (0) ]\rangle =
-\frac{\lambda^2}{2} \int_{0}^{\beta} d\tau \\
& & \langle T [
S_z (\tau )  \left( \partial_x \phi_s(\tau)- \frac{\sqrt{2}h}{v_F}
\right)
S_z (0) \left( \partial_x \phi_s (0) - \frac{\sqrt{2}h}{v_F}
\right)]\rangle,
\end{eqnarray*}
$\beta$ being the inverse temperature $T$.
Since $\phi_s$ and $S_z$ are independent, the above correlation
function
factorizes, and, by substituting the imaginary time representation of the
correlation function (\ref{SzSz}), we obtain
\begin{eqnarray*}
\delta F^{(\lambda)}_{imp}(T,h) &=& -\frac{(\lambda C)^2}{2} \int_{0}^{\beta}
d\tau
\left[
\frac{\pi/\beta}{\sin{(\pi\tau/\beta)}}\right]^{2x} \\
& &\left[ \frac{2h^2}{v_F}
+ \frac{1}{v_F^2}\left[
\frac{\pi/\beta}{\sin{(\pi\tau/\beta)}}\right]^2
\right].
\end{eqnarray*}
The term proportional to $h^2$, which defines the
magnetic susceptibility, is convergent ($x=1/3$).
The other
term, whose temperature dependence relates to the specific heat,
is instead singular, and needs to be regularized by introducing an
ultraviolet cut--off. The cut--off dependent part
is analytic in temperature (actually it is a constant apart from
$O(T^2)$ terms), so that the regularization procedure does not affect
the leading non--analytic $T$ dependence of the free energy. The result is
\begin{eqnarray*}
\delta F^{(\lambda)}_{imp}(T,h) &-& \delta F^{(\lambda)}_{imp}(0,0) =
-\lambda^2 A
\left[ T^{2x+1} + \right. \\
& & \left. \frac{2x+1}{\pi^2 x} h^2 T^{2x-1} \right]
+ O(T^2)\;,
\end{eqnarray*}
where the coefficient
\[
A=
-\frac{\pi C^2}{v_F^2}
\left( \frac{\pi}{2} \right)^{2x}
\frac{x}{2x+1}
\frac{\Gamma^2(1/2 -x)}{\Gamma(1-2x)}\;.
\]
Analogously the contribution to the free energy proportional
to $\tilde{\lambda}^2$ is:
\[
\delta F^{(\tilde{\lambda})}_{imp}(T,h)-
\delta F^{(\tilde{\lambda})}_{imp}(0,0) = \tilde{\lambda}^2 A T^{2x+1}\;.
\]
By substituting the value $x=1/3$ into the above expressions,
we get the impurity
contribution to the specific heat and susceptibility:
\[
C_{imp} \propto  T^{2/3} \; , \;\;\;
\chi_{imp} \propto T^{-1/3},
\]
in agreement with exact critical exponents found by different methods
\cite{Bethe,Tsvelik,AL}.

Another quantity of physical interest is the Wilson ratio, which
can also be evaluated, and for which we find
\[
R_W=\frac{T\chi_{imp}/C_{imp}}{T\chi_0/C_0}=
\frac{12\lambda^2}{\lambda^2+\tilde{\lambda}^2}\;,
\]
where the bulk value is taken as $T\chi_0/C_0=3/(4\pi^2)$.
It is worthy to note that, as follows from our solution,
the Wilson ratio is not universal (in spite of the spin degrees of
freedom have been kept decoupled from the charge and flavor ones).
Since $\lambda$ and $\tilde{\lambda}$ describe how one moves
away from the fixed point in the $I_z$ and $I_\perp$ directions
respectively, the non universality of $R_W$ simply reflects the
spin anisotropy. In the spin isotropic case the value $R_W=8$ was found in
Ref.~\cite{AL}.
The relevance of the spin anisotropy in
the four channel model is a consequence of the operators
$\delta H$ and $O$ having the same dimension. This contrasts
to what happens in the two channel model, where only
the former operator contributes, and in the conventional
single channel Kondo model\cite{bossbis}; in both cases
spin anisotropy does not affect the Wilson ratio.

To conclude, we have shown that the Toulouse limit
can be realized in the four channel Kondo model or, equivalently, in
the model where spin 1 conduction electrons couple to a spin 1/2 impurity.

We are thankful to A. M. Sengupta for helpful discussions and
important suggestions.


\begin{thebibliography}{99}
\bibitem{boss} P. Nozi\`eres and A. Blandin, J. Phys. (Paris) {\bf 41},
193 (1980).
\bibitem{Bethe} N. Andrei and C. Destri, Phys. Rev. Lett. {\bf 52},
364 (1984); A. M. Tsvelik and P. B. Wiegmann, Z. Phys. B {\bf 54}, 201 (1984).
\bibitem{LargeN} J. Gan, N. Andrei and P. Coleman, Phys. Rev. Lett.
{\bf 70}, 686 (1993); D. L. Cox and A. E. Ruckenstein, Phys. Rev. Lett.
{\bf 71}, 1613 (1993).
\bibitem{Tsvelik} A. M. Tsvelik, J. Phys. C {\bf 2}, 2833 (1990).
\bibitem{AL}
I. Affleck and A. W. W. Ludwig, Nucl. Phys. B, {\bf 360}, 641 (1991).
\bibitem{EK} V. J. Emery and S. Kivelson, Phys. Rev. B, {\bf 47}, 10812 (1992);
see also
D. G. Clarke, T. Giamarchi and B. I. Shraiman, {\em ibid.} {\bf 48},
7070 (1993);
A. M. Sengupta and A. Georges, {\em ibid.} {\bf 49}, 10020 (1994).
\bibitem{Toulouse} G. Toulouse, Phys. Rev. B {\bf 2}, 270 (1970).
\bibitem{G} P. Ginsparg, in {\em Fields, Strings, and Critical Phenomena},
Ed. E. Br\'{e}zin and J. Zinn-Justin, North Holland, Amsterdam, 1990.
\bibitem{noteA}
{}From the conformal field theory point of view,
the possibility of formulating the 4-channel Kondo
problem in terms of two Bose fields, Eq.(\protect\ref{hambose}\protect ),
is related to the fact that the central charge of $SU(2)_4$, $c=2$,
is equal to that one of two Bose fields
\protect\cite{G}\protect . Another candidate for a simple abelian formulation
is the 10-channel Kondo problem with $c=5/2$.
\bibitem{KF}
C. L. Kane and M. P. A. Fisher, Phys. Rev. Lett., {\bf 68}, 1220 (1992).
\bibitem{Xray}
A. O. Gogolin, Phys. Rev. Lett. {\bf 71}, 2995 (1993);
N. V. Prokof'ev, Phys. Rev. B {\bf 49}, 2243 (1994),
C. L. Kane, K. A. Matveev and L. I. Glazman, {\it ibid.} 2253 (1994).
\bibitem{schon} See K. Sch\"{o}nhammer, Prog. Theor. Phys. {\bf 106},
147 (1991), and references therein.
\bibitem{notab} This is the real current which couples to an electromagnetic
field, not to be confused with the current densities we defined
throughout the text.
\bibitem{GP} A. O. Gogolin and N. V.
Prokof'ev, Phys. Rev. B {\bf 50}, in press (1994).
\bibitem{bossbis} P. Nozi\`eres, J. Low Temp. Phys. {\bf 17}, 31 (1974).
\end{thebibliography}
\end{document}